\documentclass[%
 reprint,
 amsmath,amssymb,
 aps,
 pra,
]{revtex4-2}

\usepackage{graphicx}% Include figure files
\usepackage{dcolumn}% Align table columns on decimal point
\usepackage{bm}% bold math
\usepackage[colorlinks=true, citecolor=blue, linkcolor=blue, urlcolor=blue]{hyperref}% add hypertext capabilities
\usepackage{physics}
\usepackage[whole]{bxcjkjatype}
\usepackage[normalem]{ulem}
\usepackage{color}
\usepackage{soul}
\definecolor{electricpurple}{rgb}{0.75, 0.0, 1.0}
\definecolor{darkpastelgreen}{rgb}{0.01, 0.75, 0.24}
\definecolor{darkspringgreen}{rgb}{0.09, 0.45, 0.27}
\definecolor{forestgreen(web)}{rgb}{0.13, 0.55, 0.13}
\definecolor{green(ncs)}{rgb}{0.0, 0.62, 0.42}
\definecolor{americanrose}{rgb}{1.0, 0.01, 0.24}

\begin{document}

\preprint{APS/123-QED}

\title{Feedback-enhanced quantum reservoir computing with weak measurements}

\author{Tomoya Monomi}
\email{mono.tomo46x@gmail.com}
\author{Wataru Setoyama}
\email{setoyama@biom.t.u-tokyo.ac.jp}
\author{Yoshihiko Hasegawa}
\email{hasegawa@biom.t.u-tokyo.ac.jp}
\affiliation{Department of Information and Communication Engineering, \\ Graduate School of Information Science and Technology, \\ The University of Tokyo, Tokyo 113-8656, Japan}

\date{\today}

\begin{abstract}
Quantum reservoir computing (QRC) leverages the natural dynamics of quantum systems to process time-series data efficiently, offering a promising approach for near-term quantum devices. Unlike classical reservoir computing, the efficacy of feedback in QRC has not yet been thoroughly explored. Here, we develop a feedback-enhanced QRC framework with weak measurements. Weak measurements preserve information stored in quantum coherence, while feedback enhances nonlinearity and memory
capacity. The implementation of our framework assumes an ensemble quantum system, such as nuclear magnetic resonance. Through linear memory and nonlinear forecasting tasks, we show that our model outperforms conventional QRC approaches in many cases. Our proposed protocol achieves superior performance in systems with small measurement errors and low environmental noise. Furthermore, we theoretically demonstrate that feedback of measurement results reinforces the nonlinearity of the reservoir. These findings highlight the potential of feedback-enhanced QRC for next-generation quantum machine learning applications.
\end{abstract}

\maketitle

\section{\label{sec:introduction}INTRODUCTION}
Machine learning models based on deep neural networks have already achieved remarkable success in tasks such as image recognition and natural language processing~\cite{goodfellow2016deep,he2016deep,goodfellow2020generative,mikolov2013efficient,devlin2019bert}. However, these models require vast amounts of training data and substantial computational resources. For instance, state-of-the-art large language models contain billions to trillions of parameters~\cite{achiam2023gpt,dubey2024llama}, necessitating high-performance GPUs and significant energy consumption for both training and inference. Reservoir computing (RC) has emerged as an efficient alternative for processing time-series data~\cite{jaeger2001echo,jaeger2004harnessing,maass2002real,tanaka2019recent,nakajima2021reservoir}. RC offers fast learning and low training costs, making it an attractive approach for real-time applications. When quantum systems are employed as physical reservoir, the approach is referred to as quantum reservoir computing (QRC)~\cite{mujal2021opportunities}. Unlike conventional quantum machine learning algorithms that often require fault-tolerant quantum computers, QRC harnesses the natural analog dynamics of quantum systems for information processing. This makes it particularly well-suited for noisy intermediate-scale quantum (NISQ) devices, which refer to quantum computers with a few hundred qubits operating without error correction. The main advantage of QRC over classical RC is that the degrees of freedom of the reservoir grow exponentially with system size. In an $N$-qubit quantum spin network, the number of reservoir nodes amounts to $4^N-1$~\cite{fujii2017harnessing}. By utilizing the high-dimensional phase space provided by these nodes, QRC has the potential to surpass the capabilities of classical RC~\cite{nakajima2019boosting,nokkala2021gaussian,govia2021quantum,ghosh2019quantum,ghosh2019quantum2,tran2021learning}. Some QRC models have already been experimentally implemented~\cite{negoro2018machine,chen2020temporal,suzuki2022natural}.

\begin{figure}[t]
    \includegraphics[width=0.9\linewidth]{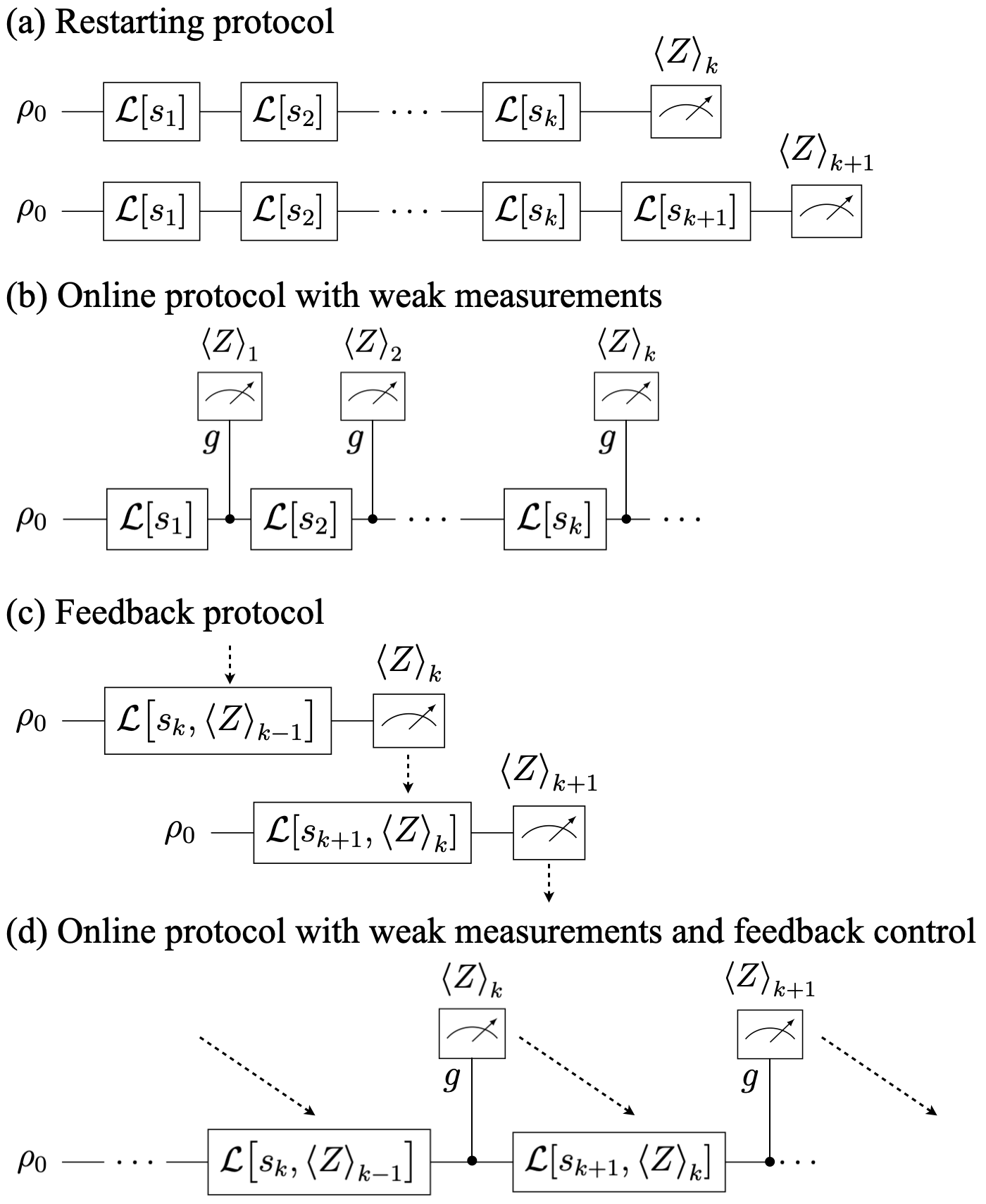}
    \caption{\label{fig:protocol} Measurement protocols in QRC with an input sequence $\{s_k\}$ and an initial state $\rho_0$. $\mathcal{L}$ represents a generalized channel parameterized by external inputs. (a) RSP. At each step, the process restarts from the first input $s_1$. (b) OLP with weak measurements. The reservoir state is continuously monitored, with measurement strength $g$. (c) FBP. The operation $\mathcal{L}$ at the $k$-th step is determined by the measurement results at the ($k-1$)-th step. The reservoir state is completely destroyed by projective measurements. (d) OLP with weak measurements and feedback control. A part of the reservoir state is carried over to the next step.}
\end{figure}
A fundamental challenge in QRC is the necessity of quantum measurement to extract information from the reservoir. Conventional QRC models rely on projective measurements, which inevitably collapse quantum states and lead to significant information loss. The most straightforward and common practice to solve this issue is the restarting protocol (RSP), which resets the system to the initial state after each measurement and restarts from the first input 
(Fig.~\ref{fig:protocol}(a)).
Although this method eliminates the effects of measurement back-action, it increases time complexity and requires external storage of previous input data. Therefore, it is not suitable for online processing. To overcome these limitations, two alternative approaches have been proposed recently~\cite{mujal2023time,kobayashi2024feedback}. One is the online protocol (OLP)~\cite{mujal2023time}
(Fig.~\ref{fig:protocol}(b)),
which employs weak measurements to continuously monitor the reservoir state while minimizing measurement back-action. By tuning the measurement strength, OLP can achieve memory performance comparable to that of RSP. The other approach is the feedback protocol (FBP), which reintegrates measurement results into the reservoir to recover information lost due to projective measurements~\cite{kobayashi2024feedback}
(Fig.~\ref{fig:protocol}(c)).
This method not only enables time-efficient computation but also offers several advantages in hardware implementation. However, despite these benefits, FBP has a critical limitation: only classical information from the previous measurement results is transmitted to the next step. Most of the information within the quantum reservoir is lost at each step, restricting access to past input data. If $N$ types of feedback values are used, the effective phase-space dimension of the reservoir cannot exceed $N$. In other words, FBP does not fully exploit the quantumness, which is the key advantage of QRC.

In this study, we propose a QRC framework that integrates weak measurements with feedback control
(Fig.~\ref{fig:protocol}(d)).
Weak measurements preserve coherence and allow the model to access more input information, while feedback enhances nonlinearity and memory capacity. To evaluate the effectiveness of this approach, we conduct numerical simulations on multiple time-series tasks, demonstrating that our model outperforms existing QRC models. Notably, our protocol is most effective in systems free from measurement errors and external noise. Furthermore, we find that feedback improves the coherence of the quantum reservoir. By analyzing the distribution of measurement results, we confirm that our model enables richer quantum dynamics than FBP. To gain deeper insight, we perform a theoretical analysis to elucidate how feedback mechanisms enhance the nonlinearity of the reservoir.

\section{\label{sec:methods}METHODS}
\subsection{Model architecture}
\begin{figure*}[t]
    \includegraphics[width=0.7\linewidth]{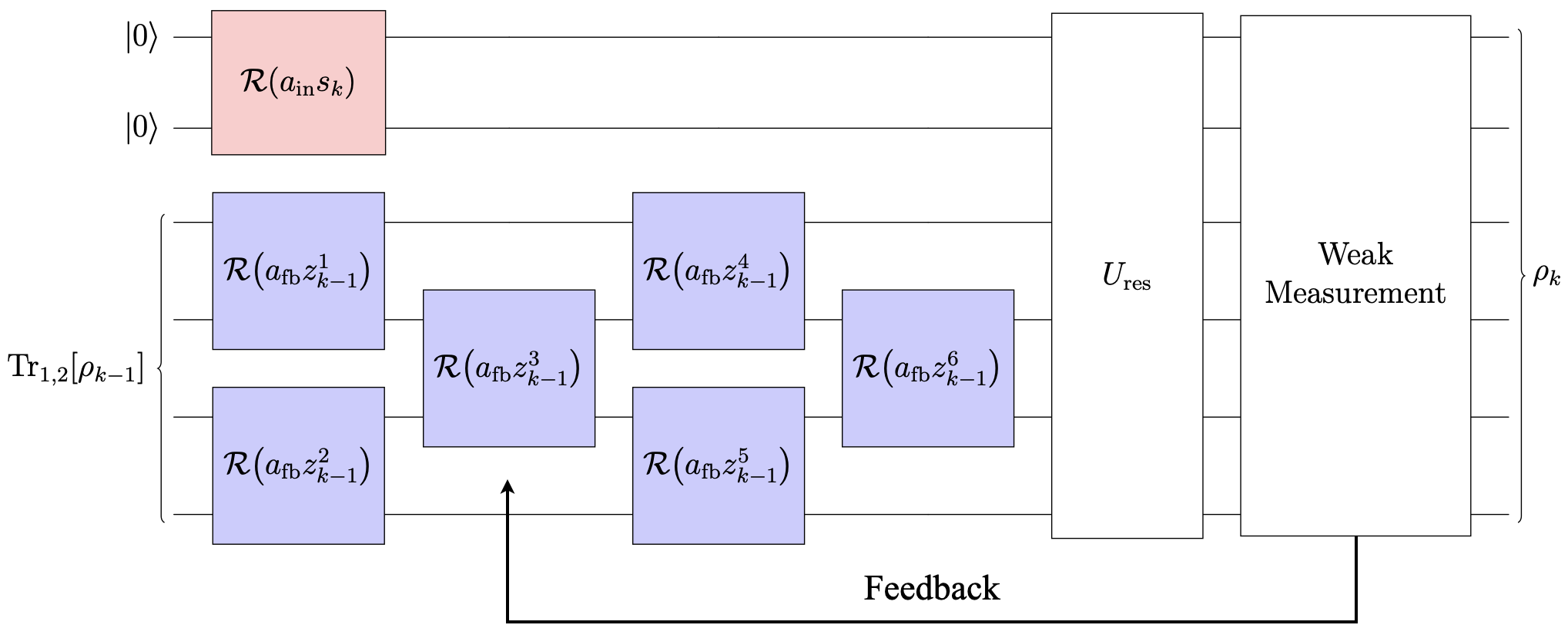}
    \caption{\label{fig:architecture} Circuit diagram of the feedback-enhanced QRC based on weak measurements at the $k$-th step. The top two qubits are used to encode the input, while the bottom four qubits store the past input information. After initializing the input qubits to $\ket{\psi_0}=\ket{00}$, the input $s_k$ is injected into qubits $1$ and $2$ via the red gate $\mathcal{R}_{1,2}(a_{\mathrm{in}}s_k)$. The previous measurement results $\mathbf{z}_{k-1}$ are fed back into qubits $3$ to $6$ via the blue gates $\mathcal{R}_{i,j}(a_{\mathrm{fb}}z_{k-1}^\alpha)$. The system then evolves under the unitary operator $U_{\mathrm{res}}$, and weak measurements are performed on all qubits.}
\end{figure*}
\begin{figure}[t]
    \includegraphics[width=0.9\linewidth]{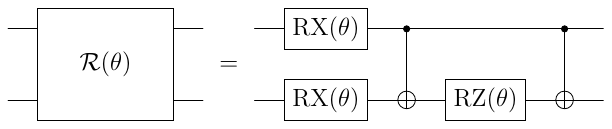}
    \caption{\label{fig:r_theta} Circuit representation of the two-qubit operator $\mathcal{R}_{i,j}(\theta)$ defined in Eq.~(\ref{eq:r_theta}).}
\end{figure}
Figure~\ref{fig:architecture}
illustrates the architecture of our feedback-enhanced QRC model with weak measurements. We set the number of spins to $N=6$. Given a one-dimensional input sequence $\{s_k\}$, we consider the task of learning a nonlinear function $y_k=f(\{s_l\}_{l=1}^k)$. The protocol consists of the following five components:
\begin{enumerate}
    \item Initialization: Input qubits are initialized to the state $\ket{\psi_0}=\ket{00}$. The density matrix of the whole system $\rho_{k-1}$ is transformed by the completely positive trace-preserving map:
    \begin{equation}
        \rho_{k-1} \to
        \ket{\psi_0}\bra{\psi_0} \otimes \mathrm{Tr}_{1,2}\,[\rho_{k-1}],
    \end{equation}
    where $\mathrm{Tr}_{1,2}\,[\cdot]$ denotes the partial trace over the input qubits, taken to be qubits $1$ and $2$, respectively.

    \item Input encoding: The input $s_k$ is injected into the reservoir via a two-qubit gate $\mathcal{R}_{1,2}(a_{\mathrm{in}}s_k)$ on qubits 1 and 2, where $a_{\mathrm{in}}$ denotes the input scaling weight. $\mathcal{R}_{i,j}(\theta)$ is defined as
    \begin{equation}
        \label{eq:r_theta}
        \mathcal{R}_{i,j}(\theta) =
        \mathrm{CX}_{i,j} \mathrm{RZ}_j(\theta) \mathrm{CX}_{i,j} \mathrm{RX}_i(\theta) \mathrm{RX}_j(\theta).
    \end{equation}
    Here, $\mathrm{CX}_{i,j}$ is the CNOT gate with control qubit $i$ and target qubit $j$, while $\mathrm{RZ}_i$ and $\mathrm{RX}_i$ are the rotation gate on qubit $i$, around the $z$- and $x$-axis, respectively. The circuit representation of $\mathcal{R}_{i,j}(\theta)$ is depicted in 
    Fig.~\ref{fig:r_theta}.
    
    \item Feedback of measurement results: The vector
    \begin{equation}
        \label{eq:z_k}
        \mathbf{z}_{k-1} = \qty[\ev{Z_1}_{k-1}, \cdots, \ev{Z_N}_{k-1}]^\top
    \end{equation}
    is constructed from the measurement results at the ($k-1$)-th step, where $Z_i$ is the Pauli $Z$ operator for qubit $i$. Each element of $\mathbf{z}_{k-1}$ is fed back into the reservoir via $\mathcal{R}_{i,j}(a_{\mathrm{fb}}z_{k-1}^\alpha)$, as shown in
    Fig.~\ref{fig:architecture},
    where $a_{\mathrm{fb}}$ denotes the feedback strength.
    
    \item Unitary evolution: An $N$-qubit gate $U_{\mathrm{res}}$ is applied to generate entanglement among all qubits. The unitary operator given by a Hamiltonian $H$ for a time interval $\Delta t$ can be written as
    \begin{equation}
        U_{\mathrm{res}} = \exp(-iH\Delta t).
    \end{equation}
    In the numerical simulations, we employ a fully connected transverse-field Ising model, which is commonly used in QRC models~\cite{fujii2017harnessing,nakajima2019boosting,tran2021learning,mujal2023time}:
    \begin{equation}
        H = \sum_{i<j} J_{ij}X_iX_j + h\sum_i Z_i,
    \end{equation}
    where $h$ is the value of the magnetic field and $J_{ij}$ is the spin-spin coupling, randomly sampled from a uniform distribution in the interval $[-J_s/2,J_s/2]$. We fix $h=5J_s$ and $\Delta t=10/J_s$ to ensure that the reservoir operates in an appropriate dynamical regime where the system can exhibit thermalization~\cite{martinez2021dynamical}.
    
    \item Weak measurements: Weak measurements are performed on all qubits, and the expectation values of observables are extracted. We measure the observables $X_i$ and $Z_i$, obtaining the vector
    \begin{equation}
        \label{eq:readout}
        \mathbf{r}_k =
        \qty[\ev{X_1}_k, \ev{Z_1}_k, \cdots, \ev{X_N}_k, \ev{Z_N}_k]^\top,
    \end{equation}
    where $X_i$ is the Pauli $X$ operator for qubit $i$. A part of the measurement results, $\mathbf{z}_k$, as defined in Eq.~(\ref{eq:z_k}), is provided back into the reservoir through the feedback connection at the subsequent ($k+1$)-th step. In a real experiment, the choice of readout nodes may depend on the specific physical implementation of the reservoir.
\end{enumerate}

The output at the $k$-th step is formed as a linear combination of the readout nodes:
\begin{equation}
    \bar{y}_k = \mathbf{w}^\top\mathbf{r}_k + b,
\end{equation}
where $\mathbf{w}$ represents the output weights and $b$ is the bias term, both optimized by minimizing the error with respect to the target output $y_k$. The output layer is trained using Ridge regression with a regularization parameter of $\alpha=10^{-8}$ to prevent overfitting.

Our QRC model employs weak measurements instead of projective measurements, which suppress measurement back-action and allow information in the quantum state to be carried over to the next step. This ensures that the entire Hilbert space of the quantum reservoir is effectively utilized. Additionally, the feedback mechanism introduces nonlinear components of the input data into the reservoir, enhancing its capacity to approximate complex functions. We note that experimental implementations of our model require platforms capable of measuring system ensembles in a single run, such as molecular ensembles~\cite{negoro2021toward} or optical pulses~\cite{garcia2023scalable}. This requirement arises because processing at each step is not independent, unlike in FBP.

\subsection{Weak measurements}
In this study, we adopt the weak measurement framework proposed in Ref.~\cite{mujal2023time}. When measuring in the $z$ direction with strength $g$, the quantum state after the measurement is given by
\begin{equation}
    \rho' = M \odot \rho,
\end{equation}
where $\odot$ represents the element-wise product and $M$ is defined as
\begin{equation}
    M = \tilde{M}^{\otimes N}, \quad
    \tilde{M} = \begin{pmatrix}
    1 & e^{-\frac{g^2}{2}} \\
    e^{-\frac{g^2}{2}} & 1
    \end{pmatrix}.
\end{equation}
The expectation value of the observable $Z_i$ is then extracted as $\mathrm{Tr}[Z_i\rho']$. To measure in the $x$ or $y$ direction, the state must be properly rotated before and after applying $M$. For the $x$ direction, we apply the Hadamard gates:
\begin{equation}
    \rho' = \mathcal{H}\qty[M\odot(\mathcal{H}\rho\mathcal{H})]\mathcal{H},
\end{equation}
where $\mathcal{H}$ is the tensor product of Hadamard gates.

In practice, the expectation values of observables are estimated by averaging over a large number of measured values. However, these estimations inherently involve statistical uncertainty. The maximum standard deviation of the mean value estimated from $N_{\mathrm{meas}}$ measurements is given by
\begin{equation}
    \label{eq:sigma1}
    \sigma = \sqrt{\frac{g^2+1}{g^2N_{\mathrm{meas}}}}
\end{equation}
for single-qubit observables $\ev{X_i},\ev{Y_i},\ev{Z_i}$, and
\begin{equation}
    \label{eq:sigma2}
    \sigma = \sqrt{\frac{g^4+2g^2+1}{g^4N_{\mathrm{meas}}}}
\end{equation}
for two-qubit observables $\ev{X_iX_j},\ev{Y_iY_j},\ev{Z_iZ_j}$ (see Supplementary of Ref.~\cite{mujal2023time}). Equation~(\ref{eq:sigma1}) and (\ref{eq:sigma2}) indicate that stronger measurements reduce statistical uncertainty, while the information stored in quantum coherence is increasingly erased.

\section{\label{sec:evaluation}PERFORMANCE EVALUATION}
In this section, we numerically evaluate the computational performance of our QRC model, which combines weak measurements and feedback control. We compare the performance with that of the FBP using projective measurements and the conventional OLP without feedback, corresponding to the case where $a_{\mathrm{fb}}=0$. The reservoir dynamics and learning procedure remain identical across all protocols. We investigate standard benchmark tasks widely used in RC models~\cite{fujii2017harnessing,nakajima2019boosting,chen2020temporal,suzuki2022natural,mujal2023time,kobayashi2024feedback,martinez2021dynamical}. Each dataset consists of $1000$ time steps. The first $20$ steps serve as a washout phase to synchronize the reservoir state with the input sequence, the next $735$ steps are used for training, and the final $245$ time steps are used for evaluation. We have verified that the washout period is sufficiently long to guarantee independence from initial conditions.

\subsection{Ideal case}
We first examine the computational performance under ideal conditions, where the number of measurements satisfies $N_{\mathrm{meas}}\to\infty$. Under this assumption, statistical uncertainty in measurements vanishes, and the average of measured values corresponds exactly to the ideal expectation value. Therefore, the performance is determined solely by the inherent characteristics of the quantum reservoir.

To assess the linear memory capacity of the reservoir, we investigate the short-term memory (STM) task. Given an input sequence $\{s_k\}$ randomly sampled from a uniform distribution in the interval $[0,1]$, the task is to reproduce the past input with a delay $\tau$:
\begin{equation}
    y_k = s_{k-\tau}.
\end{equation}
The accuracy of the output $\bar{\mathbf{y}}$ relative to the target output $\mathbf{y}$ is quantified using the coefficient of determination
\begin{equation}
    C = \frac{\mathrm{cov}^2(\mathbf{y},\bar{\mathbf{y}})}
    {\sigma^2(\mathbf{y})\sigma^2(\bar{\mathbf{y}})},
\end{equation}
where $\mathrm{cov}(\cdot,\cdot)$ and $\sigma^2(\cdot)$ denote covariance and variance, respectively. The total capacity is defined as
\begin{equation}
    C_{\Sigma} = \sum_{\tau=0}^{\tau_{\mathrm{max}}} C(\tau),
\end{equation}
where we set $\tau_{\mathrm{max}}=20$. Both $C(\tau)$ and $C_{\Sigma}$ are averaged over $20$ samples with respect to the spin-spin coupling $J_{ij}$ in the Hamiltonian $H$.

\begin{figure}[t]
    \includegraphics[width=0.9\linewidth]{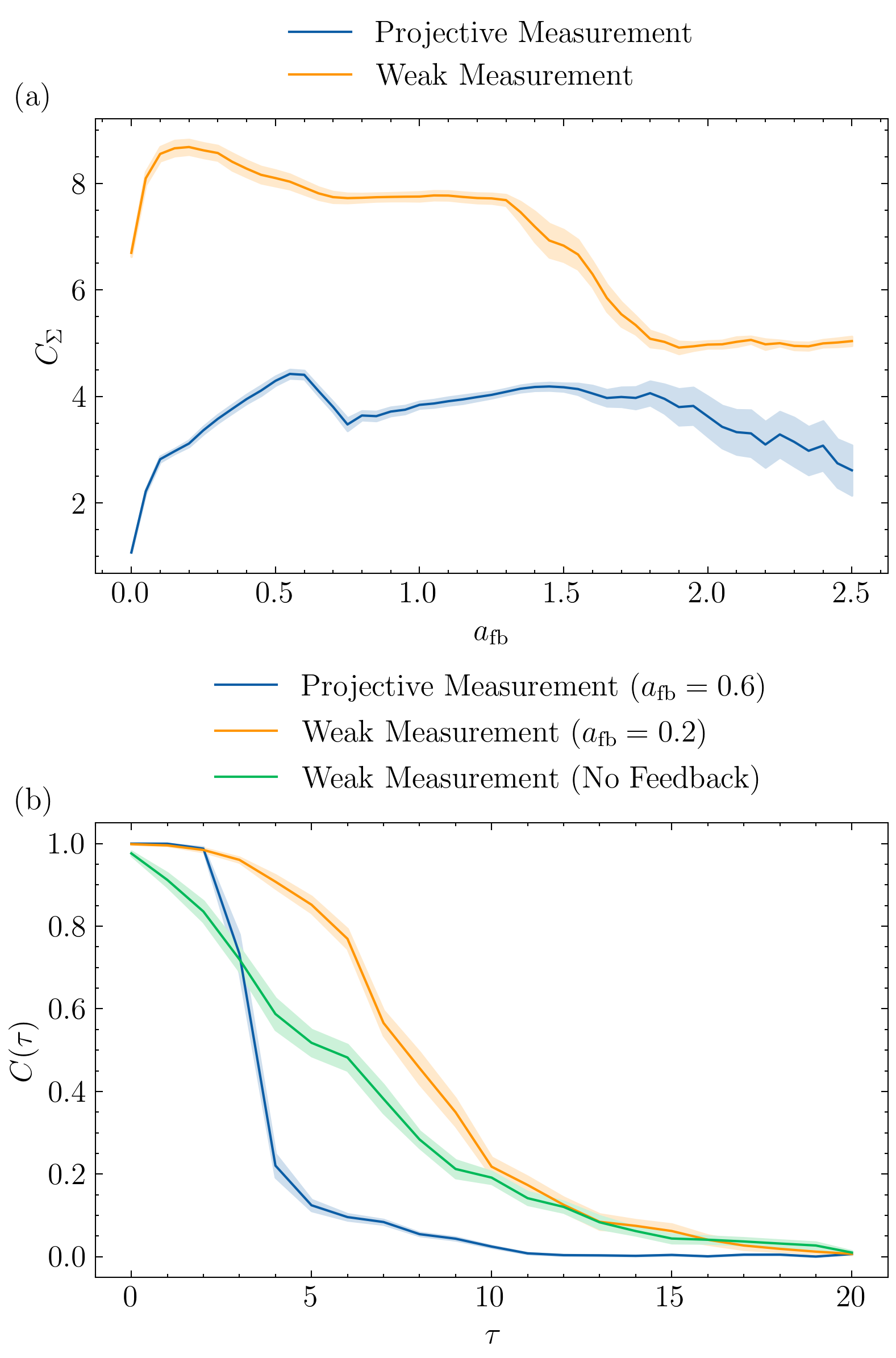}
    \caption{\label{fig:stm_ideal} STM task performance under ideal conditions. The weak measurement strength is set to $g=0.3$. (a) Total capacity $C_{\Sigma}$ plotted as a function of the feedback strength $a_{\mathrm{fb}}$. The blue and orange lines represent the results for the projective measurement-based model (FBP) and the weak measurement-based model (OLP), respectively. The shadows are the standard deviation over $20$ realizations. (b) Memory accuracy $C(\tau)$ plotted as a function of the delay $\tau$. The blue line indicates the projective measurement-based model with $a_{\mathrm{fb}}=0.6$. The orange and green lines correspond to the weak measurement-based model with $a_{\mathrm{fb}}=0.2$ and without feedback, respectively.}
\end{figure}
Figure~\ref{fig:stm_ideal}(a)
illustrates the total capacity $C_{\Sigma}$ as a function of the feedback strength $a_{\mathrm{fb}}$. The blue and orange lines denote the results for the projective measurement-based model (FBP) and the weak measurement-based model (OLP), respectively. In all subsequent experiments, the input weight is fixed at $a_{\mathrm{in}}=0.1$. As shown by the orange line, feedback in the range $0.1\leq a_{\mathrm{fb}}\leq 0.3$ significantly enhances the memory capacity of the weak measurement-based model. Notably, the maximum memory capacity achieved by this model surpasses that of the projective measurement-based model, highlighting its superior memory retention capabilities. This improvement arises from the ability of weak measurement-based QRC to retain past input information within the quantum state across time steps. However, when $a_{\mathrm{fb}}$ becomes excessively large, $C_{\Sigma}$ decreases in both models. This is because an overly strong feedback term dominates the system dynamics, disrupting its function as a reservoir.
Figure~\ref{fig:stm_ideal}(b)
presents the memory accuracy $C(\tau)$ as a function of the delay $\tau$ for fixed feedback strengths. The projective measurement-based model 
(blue line) exhibits a sharp decline for $\tau\geq4$, indicating that feedback primarily encodes information from only recent inputs. Consequently, this model is unsuitable for tasks requiring long-term memory. In contrast, the weak measurement-based model (orange and green lines) demonstrates a more gradual decay in $C(\tau)$, confirming its capacity to store and recall information over extended periods. Additionally, appropriately tuned feedback further enhances memory retention.

The second task is the nonlinear auto-regressive moving average (NARMA) task, a standard benchmark for evaluating both memory and nonlinearity in RC models~\cite{atiya2000new}. The target output for an order $n$ is expressed as
\begin{equation}
    y_k = \alpha y_{k-1} + \beta y_{k-1}\qty(\sum_{j=1}^{n} y_{k-j}) +
    \gamma s_{k-n}s_{k-1} + \delta,
\end{equation}
where $(\alpha,\beta,\gamma,\delta)=(0.3,0.05,1.5,0.1)$. We consider cases with $n=5,10,15,20$, referred to as NARMA$n$. The input sequence is generated as a superposition of sine waves:
\begin{equation}
    s_k = 0.1\qty[\sin\qty(\frac{2\pi\alpha k}{T}) \sin\qty(\frac{2\pi\beta k}{T}) \sin\qty(\frac{2\pi\gamma k}{T}) + 1],
\end{equation}
where $(\alpha,\beta,\gamma,T)=(2.11,3.73,4.11,100)$. Since the input $s_k$ is in the range $[0,0.2]$ to prevent divergences, it is rescaled to $[0,1]$ when projected into the reservoir. The prediction accuracy is measured using the normalized mean squared error (NMSE):
\begin{equation}
    \mathrm{NMSE} = \frac{\sum_k \qty(y_k - \bar{y}_k)^2}{\sum_k y_k^2}.
\end{equation}
NMSE is also averaged over $20$ samples.

\begin{figure}[t]
    \includegraphics[width=\linewidth]{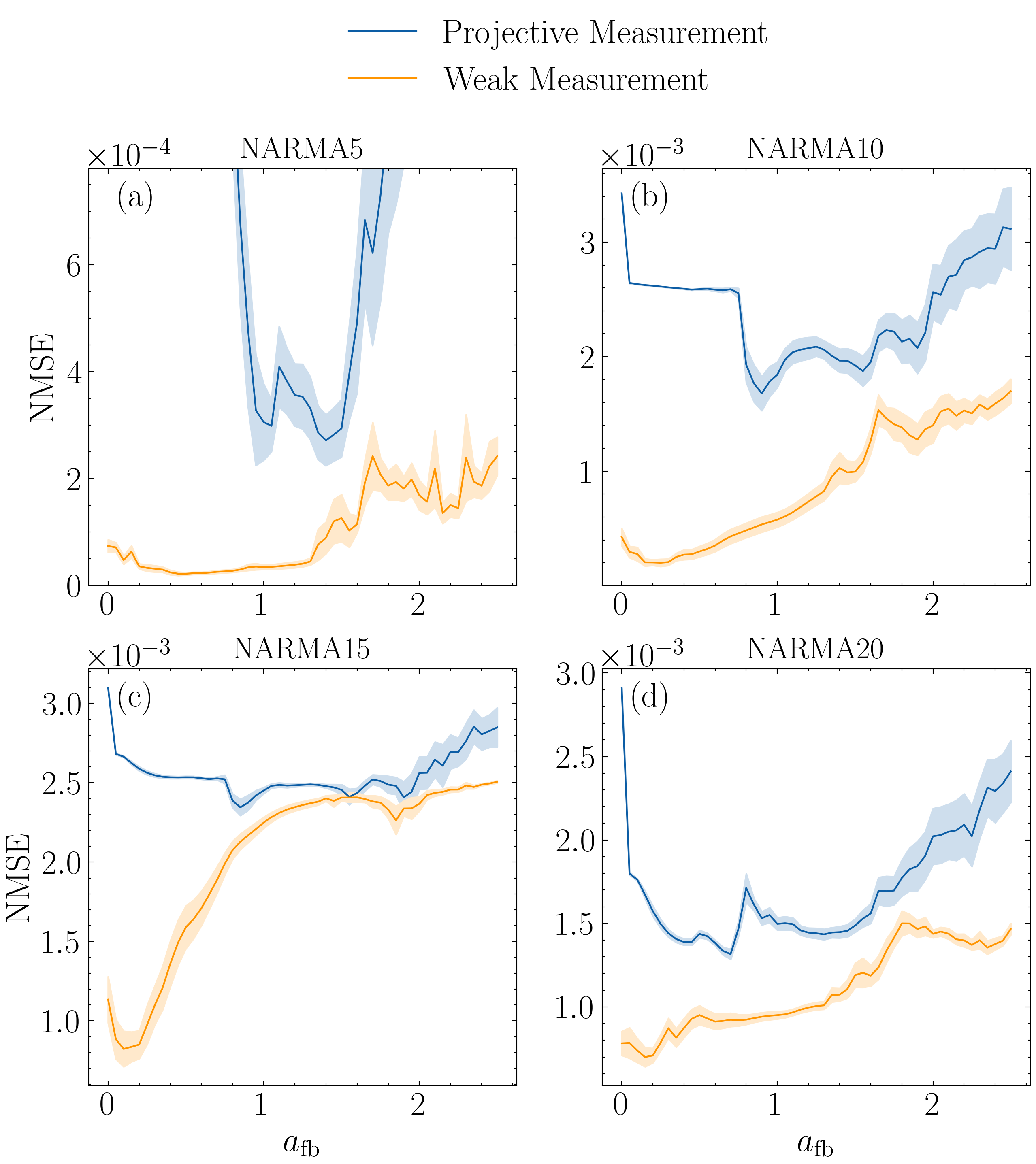}
    \caption{\label{fig:narma_ideal} NARMA task performance under ideal conditions. NMSE plotted as a function of the feedback strength $a_{\mathrm{fb}}$ for: (a) NARMA5, (b) NARMA10, (c) NARMA15, and (d) NARMA20. The weak measurement strength is set to $g=0.3$.}
\end{figure}
The results of the NARMA tasks are shown in
Fig.~\ref{fig:narma_ideal}.
In all cases, the weak measurement-based model consistently outperforms the projective measurement-based model in terms of prediction accuracy. However, for NARMA20, the improvement due to feedback is less pronounced. This suggests that while feedback reinforces information about recent inputs, it may diminish the reservoir's ability to recall inputs from further in the past.

\subsection{Case with measurement errors}
Next, we consider a more realistic scenario where expectation values are estimated from a finite number of measurement ensembles. In this case, measurement errors become unavoidable and can be modeled as Gaussian noise added to the true expectation values. When measuring an observable $O$ with $N_{\mathrm{meas}}$ measurements, the estimated value follows the probability distribution:
\begin{equation}
    P\qty(\ev{O}_{N_{\mathrm{meas}}}) =
    \mathcal{N}\qty(\ev{O}_{\infty},\sigma^2),
\end{equation}
where $\ev{O}_{\infty}$ is the ideal expectation value, and $\mathcal{N}(\mu,\sigma^2)$ denotes a normal distribution with mean $\mu$ and variance $\sigma^2$. We employ the standard deviation given by Eq.~(\ref{eq:sigma1}).

\begin{figure}[t]
    \includegraphics[width=0.9\linewidth]{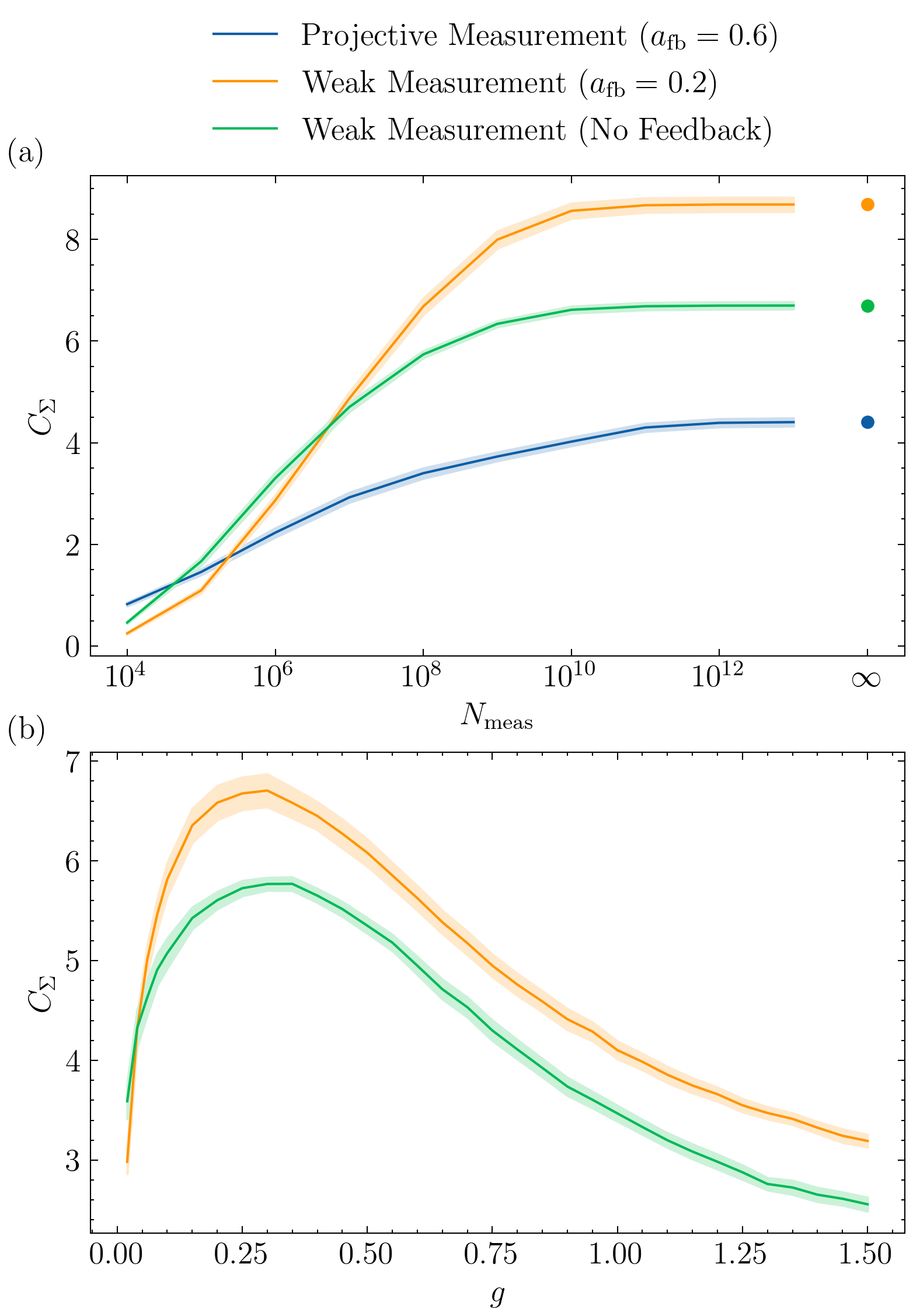}
    \caption{\label{fig:stm_Nm_g} STM task performance with a finite number of measurements. (a) Total capacity $C_{\Sigma}$ plotted as a function of the number of measurements $N_{\mathrm{meas}}$. The projective measurement strength and weak measurement strength are set to $g=10$ and $g=0.3$, respectively. (b) Total capacity $C_{\Sigma}$ plotted as a function of the measurement strength $g$. The number of measurements is set to $N_{\mathrm{meas}}=10^8$.}
\end{figure}
Figure~\ref{fig:stm_Nm_g}(a)
displays the dependence of the total capacity $C_{\Sigma}$ on the number of measurements $N_{\mathrm{meas}}$. For small $N_{\mathrm{meas}}$, the weak measurement-based model without feedback (conventional OLP) achieves the highest memory capacity. In this regime, measurement errors introduce noise into the reservoir through the feedback process, degrading performance. However, as $N_{\mathrm{meas}}$ increases, measurement errors decrease, and the weak measurement-based model with feedback (our protocol) eventually outperforms the others. By comparing convergence speeds, we observe that QRC models with feedback converge more slowly. There exists an optimal measurement strength $g$ that maximizes $C_{\Sigma}$, as shown in
Fig.~\ref{fig:stm_Nm_g}(b).
A small $g$ leads to large measurement errors, while an excessively large $g$ induces strong decoherence, reducing the amount of information retained in the quantum reservoir. The optimal value of $g$ depends on the number of measurements and the set of measured observables.

\subsection{Case with quantum noise}
In practical implementations, quantum reservoirs are subject to environmental interactions, leading to the introduction of noise. To examine its effects, we consider the depolarizing channel, a common noise model where the quantum state $\rho$ transforms into a maximally mixed state with probability $\gamma$:
\begin{equation}
    \epsilon(\rho) = (1-\gamma)\rho + \gamma\frac{I}{2^N},
\end{equation}
where $\epsilon(\cdot)$ denotes the quantum channel. Since depolarizing noise commutes with any unitary evolution, we apply it before measurement in our numerical simulations.

\begin{figure}[t]
    \includegraphics[width=0.9\linewidth]{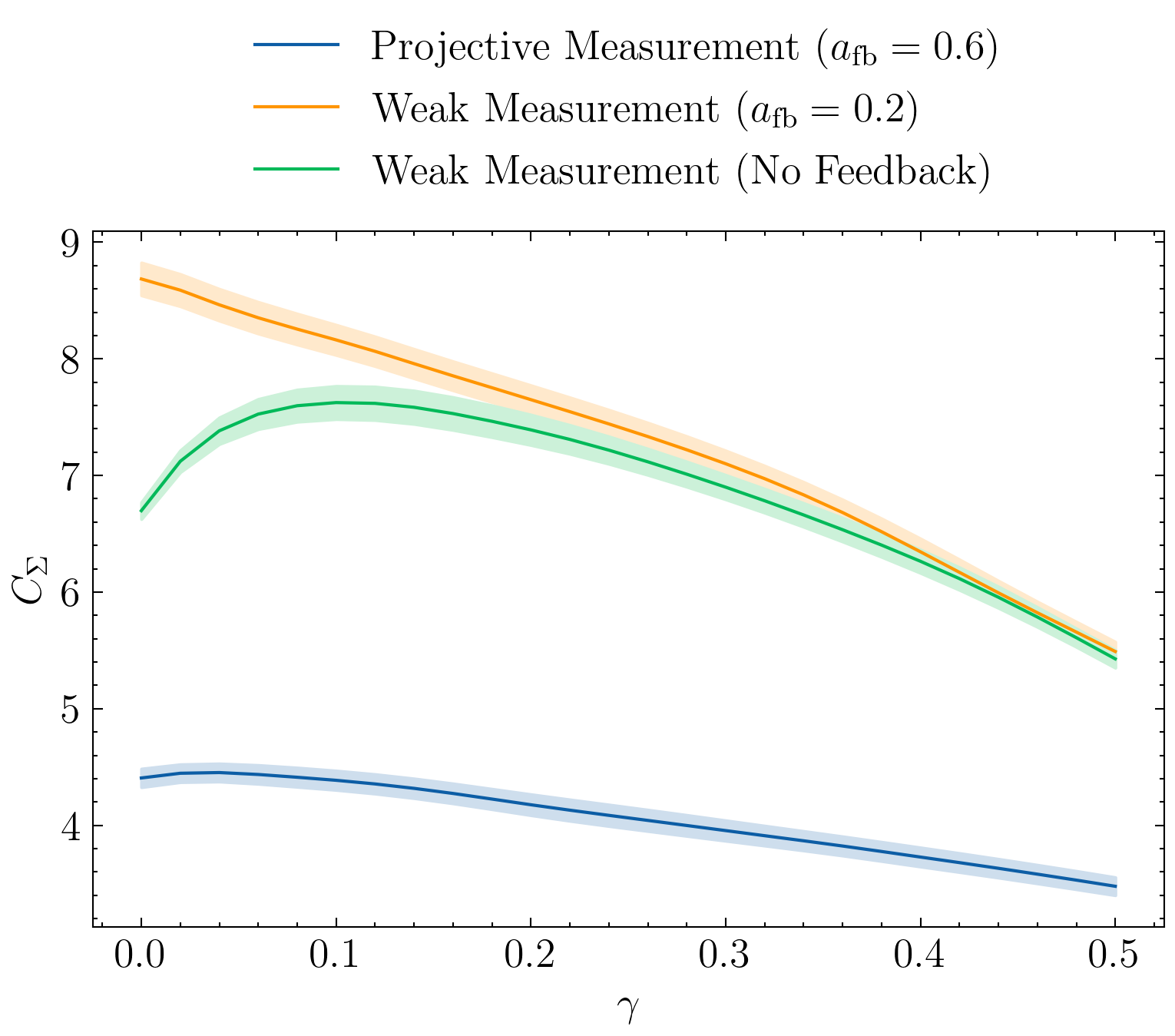}
    \caption{\label{fig:stm_noise} STM task performance under depolarizing noise. Total capacity $C_{\Sigma}$ plotted as a function of the error rate $\gamma$. The number of measurements and weak measurement strength are set to $N_{\mathrm{meas}}\to\infty$ and $g=0.3$, respectively.}
\end{figure}
The effect of depolarizing noise on the total capacity $C_{\Sigma}$ is shown in
Fig.~\ref{fig:stm_noise}.
The projective measurement-based model remains highly robust, exhibiting little degradation in memory capacity. This robustness arises because the reservoir state is reset at each step, which minimizes error propagation. In contrast, for the weak measurement-based model with feedback, $C_{\Sigma}$ monotonically decreases as the error rate $\gamma$ increases, indicating the accumulation of noise through the feedback process. Interestingly, the weak measurement-based model without feedback exhibits a region where noise slightly improves performance. This suggests that depolarizing noise may alter the fading memory properties of the reservoir, thereby enhancing its ability to distinguish the order of input sequences.

\section{\label{sec:analysis}ANALYSIS OF THE EFFECTS OF FEEDBACK}
\subsection{Coherence of the quantum reservoir}
Recent studies have highlighted the role of quantum properties, such as entanglement~\cite{gotting2023exploring} and coherence~\cite{xia2023configured,palacios2024role}, in QRC performance. Here, we focus on how feedback influences the coherence of the quantum reservoir.

The coherence of a quantum state $\rho$ is defined as
\begin{equation}
    \mathrm{QC}(\rho) = \sum_{i\neq j} \abs{\rho_{ij}},
\end{equation}
where $\rho_{ij}$ is the off-diagonal component of the density matrix. We compute the coherence during the unitary evolution governed by $U_{\mathrm{res}}$, averaging over $20$ different random realizations. The input sequence consists of random values sampled uniformly from $[0,1]$ with a length of $1000$. The first $20$ steps are discarded, and the coherence is evaluated over the remaining $980$ steps.

\begin{figure}[t]
    \includegraphics[width=0.9\linewidth]{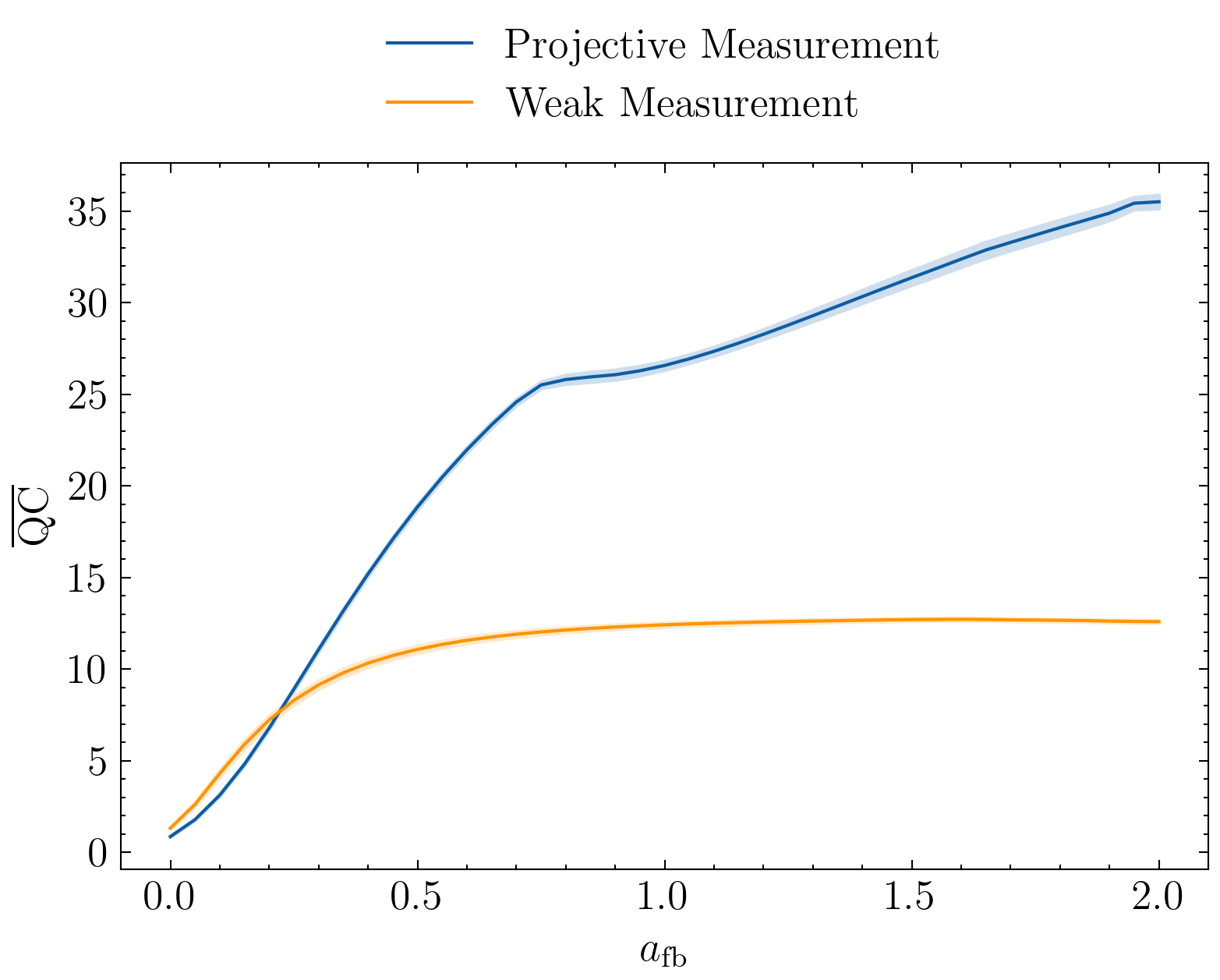}
    \caption{\label{fig:coherence} Time-averaged coherence $\overline{\mathrm{QC}}$ plotted as a function of the feedback strength $a_{\mathrm{fb}}$. The number of measurements and weak measurement strength are set to $N_{\mathrm{meas}}\to\infty$ and $g=0.3$, respectively.}
\end{figure}
In Fig.~\ref{fig:coherence},
the time-averaged coherence $\overline{\mathrm{QC}}$ is plotted as a function of the feedback strength $a_{\mathrm{fb}}$. For the projective measurement-based model, coherence increases with $a_{\mathrm{fb}}$ because the system dynamics become more complex. A similar trend is observed in the weak measurement-based model, but $\overline{\mathrm{QC}}$ saturates around $a_{\mathrm{fb}}\simeq0.7$. While previous studies suggest that coherence generally correlates with computational performance~\cite{palacios2024role}, our results indicate that coherence alone is not a sufficient condition for high performance.

\subsection{Distribution of measurement results}
One of the key advantages of QRC is the exponential scaling of phase space with system size. To analyze how effectively our QRC model utilizes this available space, we examine the distribution of measurement results. The measurement results, represented by the vector in Eq.~(\ref{eq:readout}), are projected into a lower-dimensional space using a dimensionality reduction technique called UMAP~\cite{mcinnes2018umap}. UMAP efficiently preserves both local and global structure, enabling visualization of high-dimensional data. The input sequence consists of random values sampled uniformly from $[0,1]$ with a length of $2000$. The first $20$ steps are discarded, and the remaining $1980$ measurement results are mapped into a two-dimensional space using UMAP.

\begin{figure}[t]
    \includegraphics[width=\linewidth]{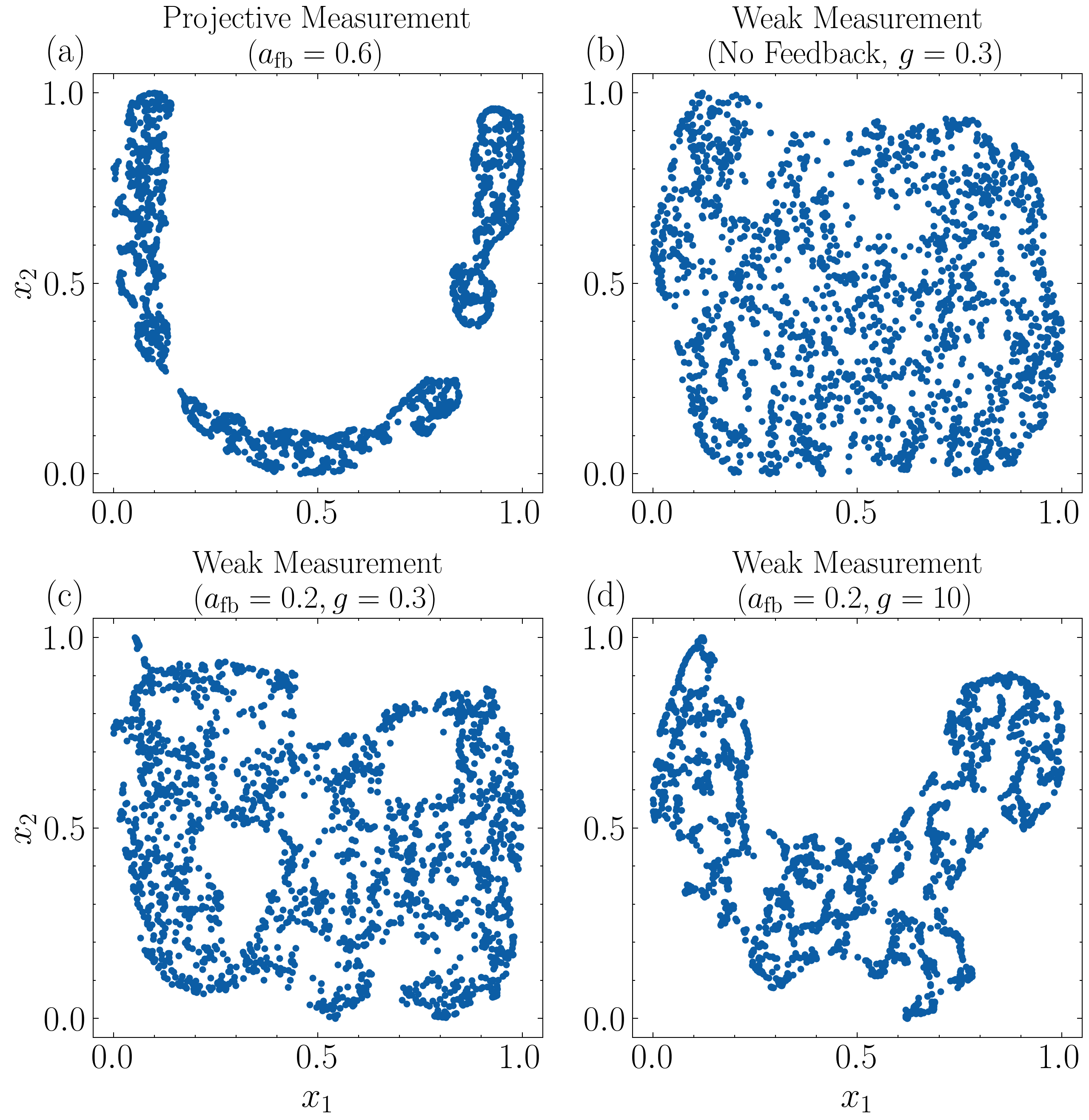}
    \caption{\label{fig:distribution} The distribution of measurement results in two-dimensional space for: (a) projective measurement-based model with $a_{\mathrm{fb}}=0.6$, (b) weak measurement-based model with $a_{\mathrm{fb}}=0,g=0.3$, (c) with $a_{\mathrm{fb}}=0.2,g=0.3$, and (d) with $a_{\mathrm{fb}}=0.2,g=10$. The variables $x_1$ and $x_2$ are selected using UMAP and scaled to the range $[0,1]$. The number of measurements is set to $N_{\mathrm{meas}}\to\infty$.}
\end{figure}
The projective measurement-based model
(Fig.~\ref{fig:distribution}(a))
and the weak measurement-based model with $g=10$
(Fig.~\ref{fig:distribution}(d))
exhibit biased distributions, suggesting that these models fail to fully explore the phase space of the quantum reservoir. This limitation likely restricts their computational capacity. In contrast, the weak measurement-based models with $g=0.3$
(Fig.~\ref{fig:distribution}(b)(c))
demonstrate more evenly spread distributions, indicating that they effectively leverage quantum coherence to maintain a rich reservoir state. These results suggest that tuning the measurement strength is essential for optimizing phase space utilization. We note that the difference between the distributions in
Fig.~\ref{fig:distribution}(b)
and
Fig.~\ref{fig:distribution}(c)
is subtle.

\subsection{Feedback-induced nonlinearity}
\begin{figure}[t]
    \includegraphics[width=0.5\linewidth]{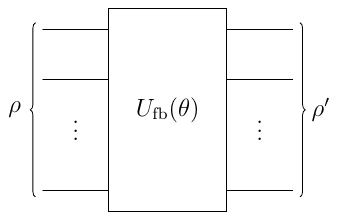}
    \caption{\label{fig:fb_circ} Simplified feedback circuit. $\rho$ and $\rho'$ represent the initial state and final state, respectively. The rotation angle $\theta$ depends on $\rho$.}
\end{figure}
To gain a deeper understanding of our QRC model, we perform a theoretical analysis of the effect of feedback on computational properties. We consider a simplified feedback circuit, as illustrated in
Fig.~\ref{fig:fb_circ},
where the feedback process consists of a single-qubit rotation:
\begin{equation}
    U_{\mathrm{fb}}(\theta) = \mathrm{RZ}_j(\theta)
    = I \otimes \cdots \mathrm{RZ}(\theta) \otimes \cdots I.
\end{equation}
Here, the rotation angle $\theta$ is determined based on the measurement results of the initial state $\rho$.

Both $\rho$ and $\rho'$ can be written as linear combinations of Pauli basis operators:
\begin{equation}
    \rho = \frac{1}{2^N}\sum_i a_iP_i, \quad a_i = \mathrm{Tr}[P_i\rho],
\end{equation}
\begin{equation}
    \rho' = \frac{1}{2^N}\sum_i b_iP_i, \quad b_i = \mathrm{Tr}[P_i\rho'],
\end{equation}
where $P_i$ is the tensor product of the Pauli operators $\{X,Y,Z,I\}$. Since the coefficients $\{a_i\}$ and $\{b_i\}$ correspond to the reservoir nodes, expanding in this basis is suitable to understand the effect of feedback. Next, we examine how the feedback gate maps the four Pauli operators to analyze the relation between $\{a_i\}$ and $\{b_i\}$. For the $\mathrm{RZ}$ gate, each Pauli operator is transformed as follows:
\begin{equation}
    X \to X\cos\theta + Y\sin\theta,
\end{equation}
\begin{equation}
    Y \to -X\sin\theta + Y\cos\theta,
\end{equation}
\begin{equation}
    Z \to Z,
\end{equation}
\begin{equation}
    I \to I.
\end{equation}
From this, we analyze the coefficient $b_i$ associated with $P_i$. For $P_i=P^1 \otimes \cdots P^{j-1} \otimes X \otimes P^{j+1} \otimes \cdots P^N$, setting $P_k=P^1 \otimes \cdots P^{j-1} \otimes Y \otimes P^{j+1} \otimes \cdots P^N$, we obtain:
\begin{align}
    b_i &= \mathrm{Tr}[P_i\rho']
    = \frac{1}{2^N} \sum_l a_l\mathrm{Tr}\qty[P_i\qty(U_{\mathrm{fb}}P_l U_{\mathrm{fb}}^\dagger)] \nonumber \\
    &= \frac{1}{2^N} \qty(a_i\mathrm{Tr}\qty[P_i\qty(U_{\mathrm{fb}}P_i U_{\mathrm{fb}}^\dagger)] + a_k\mathrm{Tr}\qty[P_i\qty(U_{\mathrm{fb}}P_k U_{\mathrm{fb}}^\dagger)]) \nonumber \\
    &= \frac{1}{2^N} \qty(a_i\cos\theta\mathrm{Tr}\qty[P_i^2] - a_k\sin\theta\mathrm{Tr}\qty[P_i^2]) \nonumber \\
    &= a_i\cos\theta - a_k\sin\theta.
\end{align}
For $P_i=P^1 \otimes \cdots P^{j-1} \otimes Y \otimes P^{j+1} \otimes \cdots P^N$, setting $P_k=P^1 \otimes \cdots P^{j-1} \otimes X \otimes P^{j+1} \otimes \cdots P^N$, we obtain:
\begin{align}
    b_i &= \mathrm{Tr}[P_i\rho']
    = \frac{1}{2^N} \sum_l a_l\mathrm{Tr}\qty[P_i\qty(U_{\mathrm{fb}}P_l U_{\mathrm{fb}}^\dagger)] \nonumber \\
    &= \frac{1}{2^N} \qty(a_i\mathrm{Tr}\qty[P_i\qty(U_{\mathrm{fb}}P_i U_{\mathrm{fb}}^\dagger)] + a_k\mathrm{Tr}\qty[P_i\qty(U_{\mathrm{fb}}P_k U_{\mathrm{fb}}^\dagger)]) \nonumber \\
    &= \frac{1}{2^N} \qty(a_i\cos\theta\mathrm{Tr}\qty[P_i^2] + a_k\sin\theta\mathrm{Tr}\qty[P_i^2]) \nonumber \\
    &= a_i\cos\theta + a_k\sin\theta.
\end{align}
For $P_i=P^1 \otimes \cdots P^{j-1} \otimes Z(I) \otimes P^{j+1} \otimes \cdots P^N$, we obtain:
\begin{align}
    b_i &= \mathrm{Tr}[P_i\rho']
    = \frac{1}{2^N} \sum_l a_l\mathrm{Tr}\qty[P_i\qty(U_{\mathrm{fb}}P_l U_{\mathrm{fb}}^\dagger)] \nonumber \\
    &= \frac{1}{2^N} a_i\mathrm{Tr}\qty[P_i\qty(U_{\mathrm{fb}}P_i U_{\mathrm{fb}}^\dagger)] \nonumber \\
    &= \frac{1}{2^N} a_i\mathrm{Tr}\qty[P_i^2] \nonumber \\
    &= a_i.
\end{align}

These results demonstrate that feedback introduces nonlinear terms with respect to $\theta$ into the reservoir nodes. Since $\theta$ depends on the initial state and, consequently, on the past input sequence, the feedback mechanism effectively integrates past information into the current reservoir dynamics, enhancing nonlinearity. This explains why feedback improves performance in the NARMA task. In the QRC model used in our numerical simulations, the feedback circuit also includes CNOT and $\mathrm{RX}$ gates. These gates introduce even more complex nonlinear transformations, enabling the reservoir to capture higher-order dependencies that would otherwise be difficult to process.

\section{\label{sec:discussion}DISCUSSION}
Compared to the conventional OLP, the key advantage of our approach is that it incorporates feedback mechanisms that enhance memory and nonlinearity. Also, our QRC model allows dynamic adjustments through the input weight $a_{\mathrm{in}}$ and feedback strength $a_{\mathrm{fb}}$, enabling flexible adaptation to diverse computational tasks. Unlike conventional physical RC, where the reservoir dynamics are fixed, our framework enables optimization of the reservoir’s internal structure through feedback connections. Furthermore, compared to the FBP using projective measurements, our approach retains information in the quantum state across time steps by employing weak measurements. The preservation of quantum coherence allows for more effective utilization of the quantum reservoir’s phase space, leading to superior memory capacity. Indeed, the analysis of measurement result distributions reveals that the weak measurement-based model explores a richer portion of the phase space, achieving greater expressive power for learning tasks.

However, our proposed protocol also has several limitations. The most significant drawback is its susceptibility to measurement errors. Since feedback relies on measurement results, errors in expectation value estimation accumulate over time, gradually degrading computational performance. This issue becomes particularly pronounced when the number of measurements is small. Additionally, our model is vulnerable to external noise. While FBP exhibits high robustness against depolarizing noise, our model suffers from performance degradation due to noise accumulation. This challenge highlights the need for error correction techniques that target depolarizing noise~\cite{fitzek2020deep}.

A promising platform for implementing our model is nuclear magnetic resonance (NMR) spin ensemble systems~\cite{negoro2018machine,negoro2021toward}, where measurement results are averaged over $10^{18}\sim10^{20}$ copies of the same molecules. This massive averaging reduces statistical uncertainty, making NMR an ideal candidate for implementation. Moreover, the measurement strength in these systems can be lowered to a level where measurement back-action becomes nearly negligible. Although non-temporal tasks have already been demonstrated using NMR, temporal data processing has yet to be realized. Exploring the feasibility of feedback-enhanced QRC in NMR systems represents an exciting direction for future research.

\section{\label{sec:conclusion}CONCLUSION}
In this study, we proposed the feedback-enhanced QRC framework based on weak measurements. This approach incorporates feedback mechanisms into the reservoir while preserving quantum coherence, thereby improving memory and nonlinearity compared to existing QRC models. We note that implementing this framework requires platforms where system ensembles can be measured at each step, such as NMR systems. To evaluate its computational performance, we conducted numerical experiments on the STM and NARMA tasks, which are commonly used benchmarks for time-series processing. Our results showed that the model outperforms existing QRC models when the feedback strength is properly tuned. Despite these advantages, the performance deteriorates in the presence of measurement errors or depolarizing noise. To further understand the role of feedback, we analyzed the effects on quantum coherence and measurement result distributions. Our findings indicated that feedback improves coherence up to a certain threshold but does not significantly affect the effective phase-space dimension. Additionally, we theoretically demonstrated that feedback mechanisms enhance nonlinearity by incorporating past measurement outcomes into the reservoir dynamics.

Several avenues remain for extending the capabilities of feedback-enhanced QRC. First, the selection of feedback observables significantly impacts performance, as discussed in Appendix~\ref{sec:appendix}. Further research is needed to determine optimal feedback strategies tailored to different task categories, such as chaotic time series prediction and high-dimensional pattern recognition. Second, incorporating quantum error mitigation techniques~\cite{cai2023quantum,strikis2021learning,takagi2022fundamental,guo2022quantum} could enhance the stability of the model. In the NISQ era, it is worth to develop adaptive feedback mechanisms that dynamically adjust to environmental noise conditions, ensuring greater robustness in practical implementations. We believe that these investigations will play a crucial role in unlocking the full potential of feedback-enhanced QRC, paving the way for more reliable and efficient quantum machine learning applications.

\begin{acknowledgments}
This work was supported by JSPS KAKENHI Grant Number JP23K24915.
\end{acknowledgments}

\appendix*
\section{\label{sec:appendix}Optimal feedback values}
In the feedback-enhanced QRC model proposed in this study, the expectation value of the obserable $Z$ used as the feedback signal to the reservoir. In this section, we analyze the effects of different feedback values on computational performance.

\begin{figure}[b]
    \includegraphics[width=0.9\linewidth]{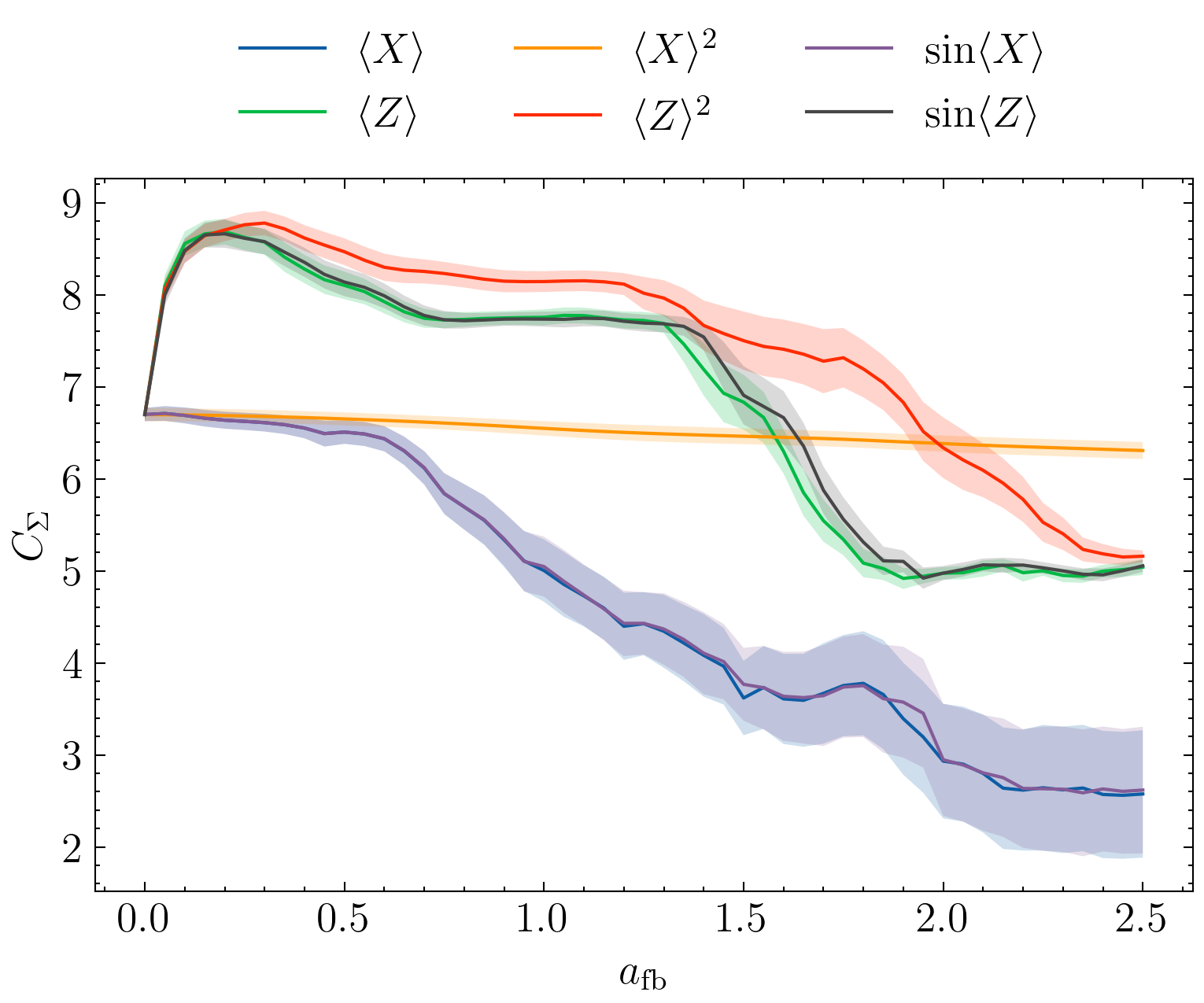}
    \caption{\label{fig:stm_ev} Total capacity $C_{\Sigma}$ plotted as a function of the feedback strength $a_{\mathrm{fb}}$ for different feedback values. The number of measurements and weak measurement strength are set to $N_{\mathrm{meas}}\to\infty$ and $g=0.3$, respectively.}
\end{figure}
Figure~\ref{fig:stm_ev}
depicts the total capacity $C_{\Sigma}$ as a function of the feedback strength $a_{\mathrm{fb}}$ for different feedback values. Memory capacity improves when $\ev{Z},\ev{Z}^2$, and $\sin\ev{Z}$ are used as feedback values, whereas it deteriorates when $\ev{X},\ev{X}^2$, and $\sin\ev{X}$ are used. This suggests that information related to past inputs is more effectively encoded in the $z$ basis. The superiority of $\ev{Z}$-based feedback may be attributed to the fact that qubits in the transverse-field Ising model are mostly aligned along the $z$ direction~\cite{martinez2021dynamical}. The optimal choice of feedback values is likely to depend on the inherent dynamics properties of the system.

\bibliography{apssamp}

\end{document}